\documentclass[aps, prl, twocolumn, superscriptaddress]{revtex4}
\usepackage{graphicx}
\usepackage{dcolumn}
\usepackage{bm}
\usepackage{natbib}

\newcommand*{\cm}[1]{#1~cm$^{-1}$}

\begin{document}

\title{Soft-mode behavior of electromagnons in multiferroic manganite}

\author{A. M. Shuvaev}
\affiliation{Experimentelle Physik IV, Universit\"{a}t W\"{u}rzburg,
97074 W\"{u}rzburg, Germany} %
\author{J. Hemberger}
\author{D. Niermann}
\affiliation{II. Physikalisches Institut, Universit\"{a}t zu K\"{o}ln, 50937 K\"{o}ln, Germany} %
\author{F. Schrettle}
\author{A. Loidl}
\affiliation{Experimentalphysik V, EKM, Universit\"{a}t Augsburg, 86135 Augsburg, Germany} %
\author{V. Yu. Ivanov}
\author{V. D. Travkin}
\author{A. A. Mukhin}
\affiliation{General Physics Institute, Russian Acad. Sci., 119991
Moscow, Russia}
\author{A. Pimenov}
\affiliation{Experimentelle Physik IV, Universit\"{a}t W\"{u}rzburg,
97074 W\"{u}rzburg, Germany} %

\date{\today}

\begin{abstract}
The behavior of the low-frequency electromagnon in multiferroic
DyMnO$_3$ has been investigated in external magnetic fields and in a
magnetically ordered state. Significant softening of the
electromagnon frequency is observed for external magnetic fields
parallel to the a-axis ($B\|a$),  revealing a number of similarities
to a classical soft mode behavior known for ferroelectric phase
transitions. The softening of the electromagnon yields an increase
of the static dielectric permittivity which follows a similar
dependence as predicted by the Lyddane-Sachs-Teller relation. Within
the geometry $B\|b$ the increase of the electromagnon intensity does
not correspond to the softening of the eigenfrequency. In this case
the increase of the static dielectric permittivity seem to be
governed by the motion of the domain walls.
\end{abstract}


\maketitle

\section{Introduction}
Soft modes have long been studied in connection with structural
phase transitions in crystalline solids~\cite{blinc_book,
cochran_ap_1960, scott_rmp_1974}. Close to such phase transition the
frequency of the normally lowest lattice vibration strongly
decreases~\cite{petzelt_f_1987} (softens) as a result of a
"flattening" of an effective potential. Therefore, soft modes
reflect an instability of the crystal lattice~\cite{cowley_ap_1980}
and are often accompanied by a substantial nonlinearity of the
system. Especially in ferroelectric crystals the softening of a
lattice vibration is followed by a divergence of the static
dielectric permittivity. In a simple case a lattice vibration can be
written in the Lorentz form,
\begin{equation}\label{lo}
\varepsilon(\omega)=\varepsilon_{\infty}+\frac{\Delta\varepsilon
\cdot \omega_{r}^2}{\omega_{r}^2-\omega^2-i\omega \gamma}=
\varepsilon_{\infty} \frac{\omega_{LO}^2-\omega^2-i\omega
\gamma}{\omega_{r}^2-\omega^2-i\omega \gamma} \ .
\end{equation}
Here $\omega_{r}$ is the resonance frequency, $\omega_{LO}$ is the
frequency of the longitudinal phonon, $\gamma$ is the damping,
$\varepsilon_{\infty}$ is the high frequency permittivity, and
$\Delta\varepsilon$ is the dielectric contribution of the mode to
the static dielectric permittivity. From this equation the well
known Lyddane-Sachs-Teller (LST) relation immediately follows
\cite{scott_rmp_1974}
\begin{equation}\label{lst}
\varepsilon(0)/\varepsilon_{\infty}=(\varepsilon_{\infty}+\Delta
\varepsilon)/\varepsilon_{\infty}=[\omega_{LO}/\omega_{r}]^2 \ .
\end{equation}
Close to structural phase transition in ferroelectrics $\omega_r$
softens and $\omega_{LO}$ basically remains constant
\cite{blinc_book}, which leads to the mentioned divergence of the
static dielectric permittivity.

Physically similar conclusion can be drawn taking into account that
the spectral weight of a single excitation is often conserved. In
spectroscopy this is generally termed as a sum rule and can be
derived from first principles~\cite{dressel_book_2002}. The spectral
weight is proportional to the electron density and can be written as
$S\sim\Delta\varepsilon \cdot \omega_r^2$. Evidently, in order to
preserve the constant spectral weight with decreasing resonance
frequency, the dielectric contribution must diverge as
$\Delta\varepsilon\sim 1/\omega_r^{2}$. The last formula works well
for the soft lattice vibrations and represents another form of the
LST relation.

Although governed by different mechanisms, qualitatively similar
effects can be observed in glassy materials as well. In this case a
broad structural relaxation mode, called
alpha-relaxation~\cite{lunkenheimer_cp_2000}, moves towards zero
frequency reflecting the freezing of the ionic movement. Contrary to
the crystalline solids, in glasses the situation is more complicated
and the spectral weight of the structural relaxation is not
conserved. However, the divergence of the static dielectric constant
close to glass transition has been put into discussion for
non-crystalline solids as well~\cite{menon_prl_1995}.

Another sum rule can be derived in case of dielectric spectroscopy,
which of course correlates with the conservation of the spectral
weight. This sum rule states that the static dielectric permittivity
can be calculated as a sum of all contributions from absorption
processes at finite frequencies~\cite{dressel_book_2002}. Therefore,
in experiment one might try to always correlate the changes in
static properties with changes in the absorption spectra at finite
frequencies. As a characteristic example, the suppression of an
electro-active mode in multiferroic GdMnO$_3$ and TbMnO$_3$ has been
put forward as a spectroscopic explanation of field induced changes
in the static dielectric constant~\cite{pimenov_nphys_2006}.

Multiferroics represent an intriguing class of materials in which
electric and magnetic orders coexist~\cite{fiebig_jpd_2005,
tokura_science_2006, eerenstein_nature_2006}. Most interesting
effects occur if both orders are strongly coupled. This leads to
such effects like the control of electric polarization by external
magnetic field or the control of magnetization by electric field. In
addition to static polarization, dynamic properties of multiferroics
are very rich and in various compounds show the existence of a
series of new excitations. These excitations are electrically active
magnetic modes of the cycloidal spin structure and they have been
called electromagnons~\cite{pimenov_jpcm_2008, sushkov_jpcm_2008}.
The electromagnons may be suppressed in external magnetic field
leading to substantial changes of the dielectric permittivity in a
broad frequency range.

DyMnO$_3$ belongs to the most studied multiferroic manganites with
orthorhombic structure. Below the N\'{e}el phase transition at
$T=39$\,K DyMnO$_3$ first possesses an incommensurate magnetic
order~\cite{kimura_prb_2003, kimura_prb_2005, feyerherm_prb_2006}.
For  $T\leq19$\,K this order turns into a spin cycloid with the
manganese spins rotating in the crystallographic $bc$ plane. As has
been proven both theoretically and experimentally, this spin
structure leads to the occurrence of the static electric
polarization parallel to the $c$ axis~\cite{cheong_nmat_2007,
goto_prl_2004, kimura_prb_2005}. Similar to such multiferroics like
GdMnO$_3$ or TbMnO$_3$, DyMnO$_3$ shows the series of electromagnons
at finite frequencies, which basically consists of two modes at 2
and 6 meV (15 and \cm{50}), respectively~\cite{pimenov_jpcm_2008,
kida_prb_2008,kida_josab_2009}. Although the physical mechanism of
the electromagnons is not fully understood until now, the high
frequency mode seems to correspond to a zone boundary
magnon~\cite{aguilar_prl_2009, lee_prb_2009}. This magnon acquires
electric dipole activity and becomes visible in the optical spectra
as a result of the Heisenberg exchange mechanism combined with the
cycloidal spin structure. The up-to-date situation with the
low-frequency electromagnon is a bit more complicated. According to
recent experimental results~\cite{pimenov_prl_2009,
shuvaev_prl_2010} on the closely similar multiferroic TbMnO$_3$, the
low-frequency electromagnon corresponds to an eigenmode of the
cycloidal spin structure~\cite{senff_jpcm_2008} which becomes
infrared active due to an incommensurate spin modulation of the
cycloid. We note that an alternative explanation based on
anisotropic effects has been suggested as
well~\cite{stenberg_prb_2009,mochizuki_prl_2010}.

In this work we present the comparison of the electromagnon dynamics
in external magnetic fields with the changes in the static
dielectric permittivity in the multiferroic DyMnO$_3$. Substantial
decrease of the electromagnon frequency is observed in the cycloidal
magnetic phase. This softening behavior correlates with an increase
of the static dielectric permittivity, thus revealing similar
dependence as predicted by the Lyddane-Sachs-Teller relation.

\begin{figure}
\includegraphics[width=1.0\linewidth, clip]{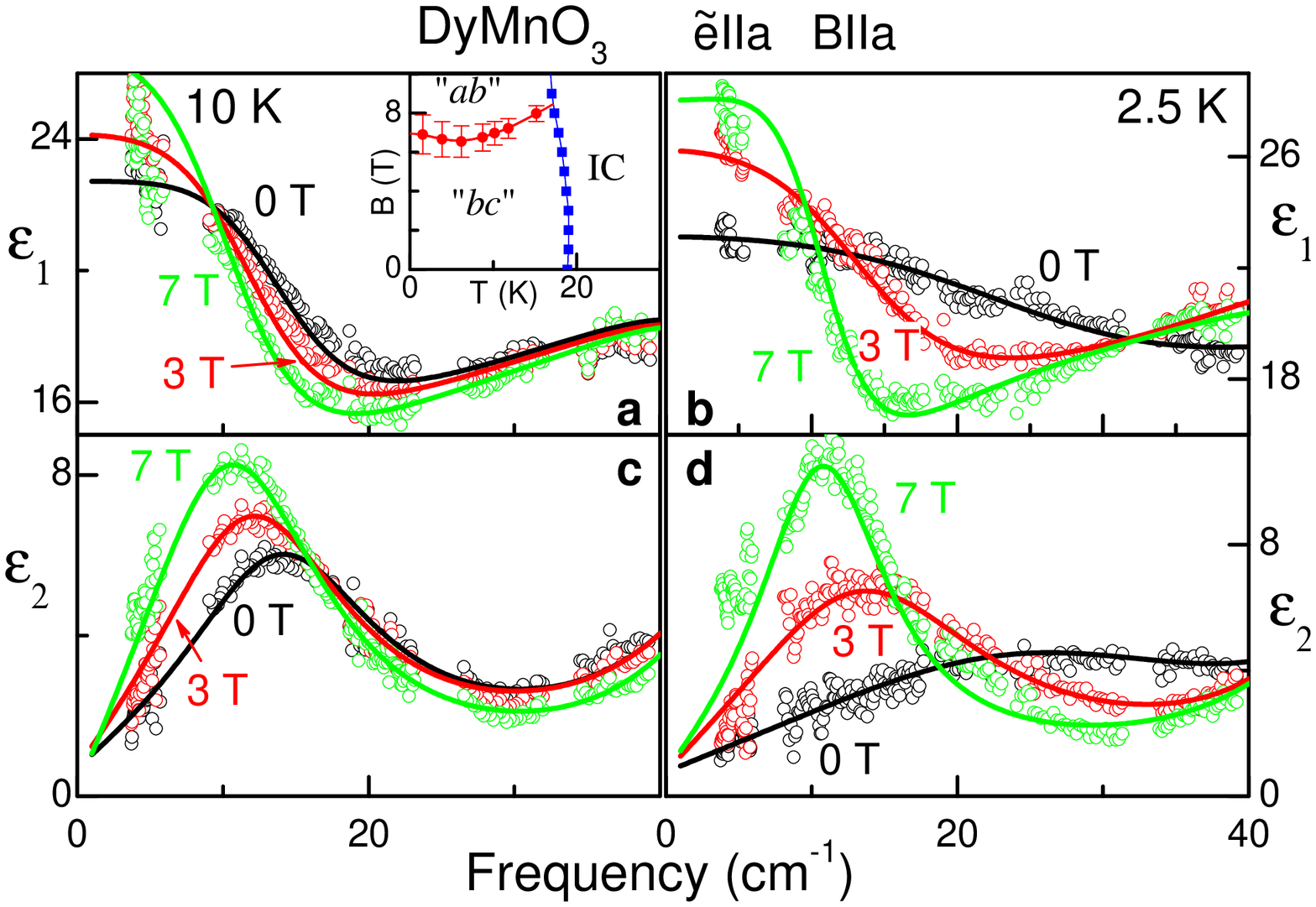}
\caption{Frequency dependence of the real (\textbf{a},\textbf{b})
and imaginary (\textbf{c},\textbf{d}) parts of the dielectric
permittivity $\varepsilon^*=\varepsilon_1+i\varepsilon_2$ of
DyMnO$_3$ along the $a$ axis for different temperatures and external
magnetic fields. For increasing magnetic fields a softening and a
growth of intensity of the electromagnon is clearly observed. The
inset shows the (B,T) phase diagram for static magnetic fields along
the $a$ axis. IC - incommensurate magnetic state, ``\textit{ab}''
and ``\textit{bc}'' denote the \textit{ab}-plane and the
\textit{bc}-plane oriented cycloids, respectively.  Error bars for
the $bc$-$ab$ transition reflect the hysteresis and history
dependence in different experiments. $\tilde{e}$ indicates the ac
electric field of the electromagnetic wave.} \label{fspectra_a}
\end{figure}

\section{Experimental details}
The spectroscopic experiments in the terahertz frequency range
(\cm{3} $< \nu <$ \cm{40}) have been carried out in a Mach-Zehnder
interferometer arrangement~\cite{volkov_infrared_1985,
pimenov_prb_2005} which allows measurements of the amplitude and the
phase shift in a geometry with controlled polarization of radiation.
Dynamic dielectric properties
$\varepsilon^*(T,B)=\varepsilon_1+i\varepsilon_2$ were calculated
from these quantities using the Fresnel optical equations for the
complex transmission coefficient. The experiments in external
magnetic fields up to 8~T have been performed in a superconducting
split-coil magnet with polypropylene windows. A frequency-response
analyzer (Novocontrol alpha-analyzer) was used for static dielectric
measurements. Single crystals of DyMnO$_3$ have been grown using the
floating-zone method with radiation heating. The samples were
characterized using X-ray, magnetic, dielectric and optical
measurements~\cite{pimenov_jpcm_2008, schrettle_prl_2009}. The
results of these experiments including the magnetic phase diagrams
are closely similar to the published results~\cite{kimura_prb_2005}.

\section{Results and discussion}
\subsection{$B\|a$}

Figure \ref{fspectra_a} shows typical spectra of the low frequency
electromagnon in DyMnO$_3$, which have been obtained in the
magnetically ordered state with the spin cycloid oriented in the
crystallographic $bc$ plane. This phase is indicated as
``\textit{bc}'' in the phase diagram shown in the inset to Fig.
\ref{fspectra_a}. Application of an external magnetic field parallel
to the crystallographic $a$ axis leads to a rotation of the $bc$
cycloid around the $b$ axis from the $bc$ plane to the $ab$ plane.
The spectra in Fig. \ref{fspectra_a} clearly demonstrate that in
external magnetic fields $B\|a$ the electromagnon shifts to lower
frequencies and gains intensity. This behavior reveals already at
this point a close similarity to classical soft modes. In order to
investigate this similarity in more details, we have carried out the
qualitative analysis of the electromagnon using the
Lorentz-oscillator model and compared these results with the static
dielectric permittivity.

\begin{figure}
\includegraphics[width=0.7\linewidth, clip]{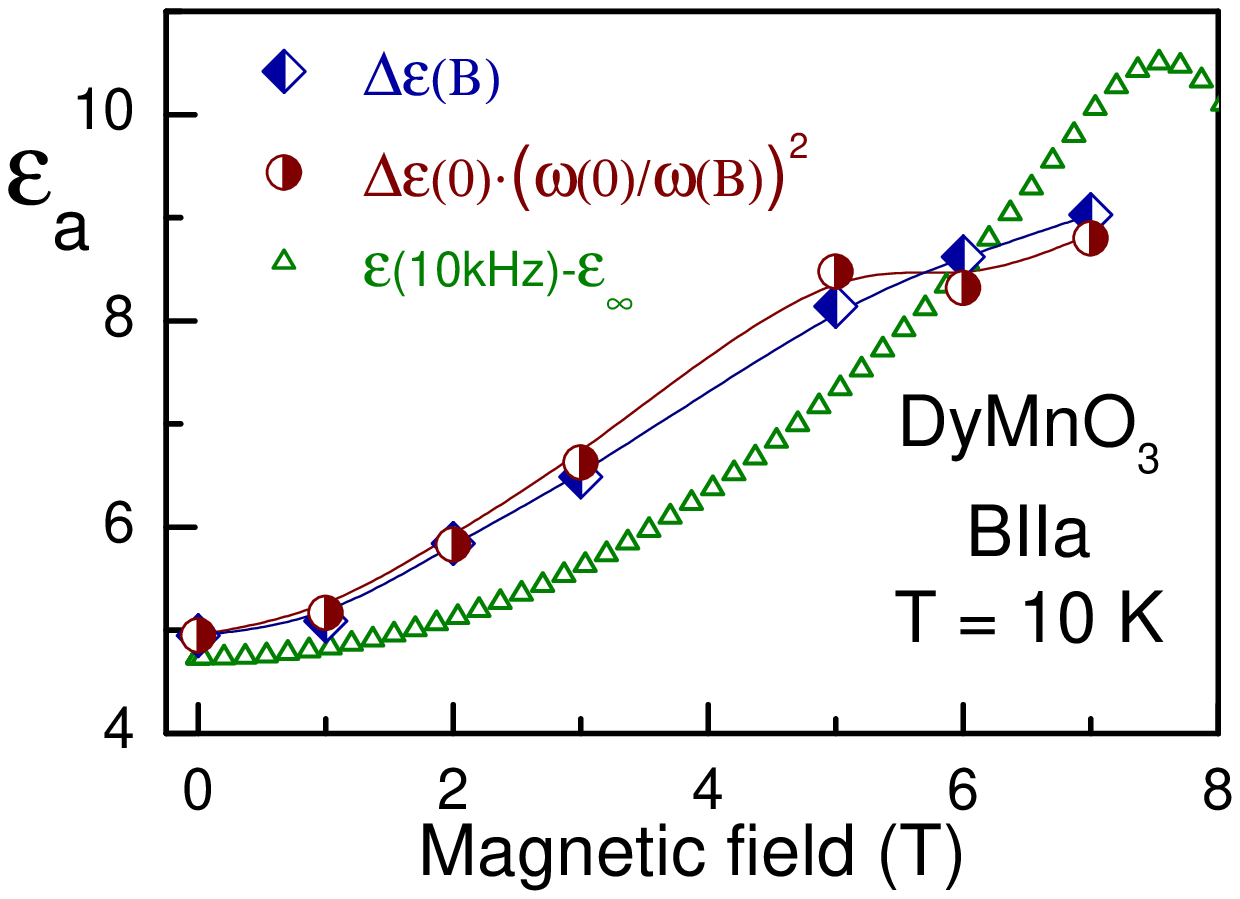}
\caption{Comparison of the static and dynamic properties of
DyMnO$_3$ along the $a$ axis. Diamonds represent the measured
dielectric contribution of the electromagnon. Circles - dielectric
contribution of the electromagnon as predicted by the LST relation.
Triangles - static dielectric permittivity with a high-frequency
value $\varepsilon_{\infty}=25$ subtracted.  } \label{fpar_a}
\end{figure}

Figure \ref{fpar_a} shows the magnetic field dependence of the
dielectric contribution of the electromagnon in comparison with the
static permittivity. Blue diamonds represent the field dependence of
the dielectric contribution $\Delta\varepsilon$ of the low-frequency
electromagnon showing an increase by more than a factor of two as
approaching the phase transition to the $ab$-oriented spiral around
7\,T. The dielectric contribution closely correlates with the
decrease of the resonance frequency, which is demonstrated by
plotting $\Delta\varepsilon\cdot\frac{\omega(0)^2}{\omega(B)^2}$
(red circles). This plot corresponds directly to the
Lyddane-Sachs-Teller relation and reflects the conservation of the
spectral weight of the electromagnon in external fields parallel to
the $a$ axis. Open triangles in Fig. \ref{fpar_a} show the field
dependence of the static dielectric constant in DyMnO$_3$ as
measured at 10 kHz. Here we subtracted the contribution from the
higher frequency processes ($\varepsilon_{\infty}=25$) which is
given by electronic transition, phonons, and a second
electromagnon~\cite{kida_prb_2008} at 6 meV. We observe a close
correlation between static and dynamic properties in spite of more
than seven orders of magnitude difference in frequency. According to
the sum rules mentioned above, this result demonstrates that for the
geometry $B\|a$ the changes in the static properties are nearly
completely governed by the softening of the electromagnon and no
other contributions exist between kHz and THz frequencies.

We note that a correspondence of lattice soft modes and
electromagnons is not straightforward. The latter are defined in the
ordered state and reveal critical behavior approaching a phase
transition as function of magnetic field. The LST-derived equation
for DyMnO$_3$ relates the dielectric constant at zero magnetic
fields to that in finite fields, while it relates transverse and
longitudinal modes in dielectrics. The softening of the
electromagnon in external fields can be qualitatively understood
taking into account the switching of the orientation of the spin
cycloid. Similar to many other structural transitions the effective
stiffness of the cycloid probably tends to zero on the phase border
between the $bc$- and $ab$ cycloids, leading to a softening of the
electromagnon. The unresolved question is: why the spectral weight
of the electromagnon is conserved during the softening of the
eigenfrequency? In case of classical softening of the lattice
vibration one normally argues that the spectral weight of the soft
mode is directly connected to the total number of electrons in the
material. In agreement with the charge conservation a constant
spectral weight may be expected for soft phonons. In case of a
magnetic cycloid the electromagnon gains the spectral weight as a
result of a complex interplay of various mechanisms. Therefore, we
cannot use the conservation of the magnetic moment as an argument,
and the observed conservation of the spectral weight remains an
actual problem.

\begin{figure}
\includegraphics[width=0.6\linewidth, clip]{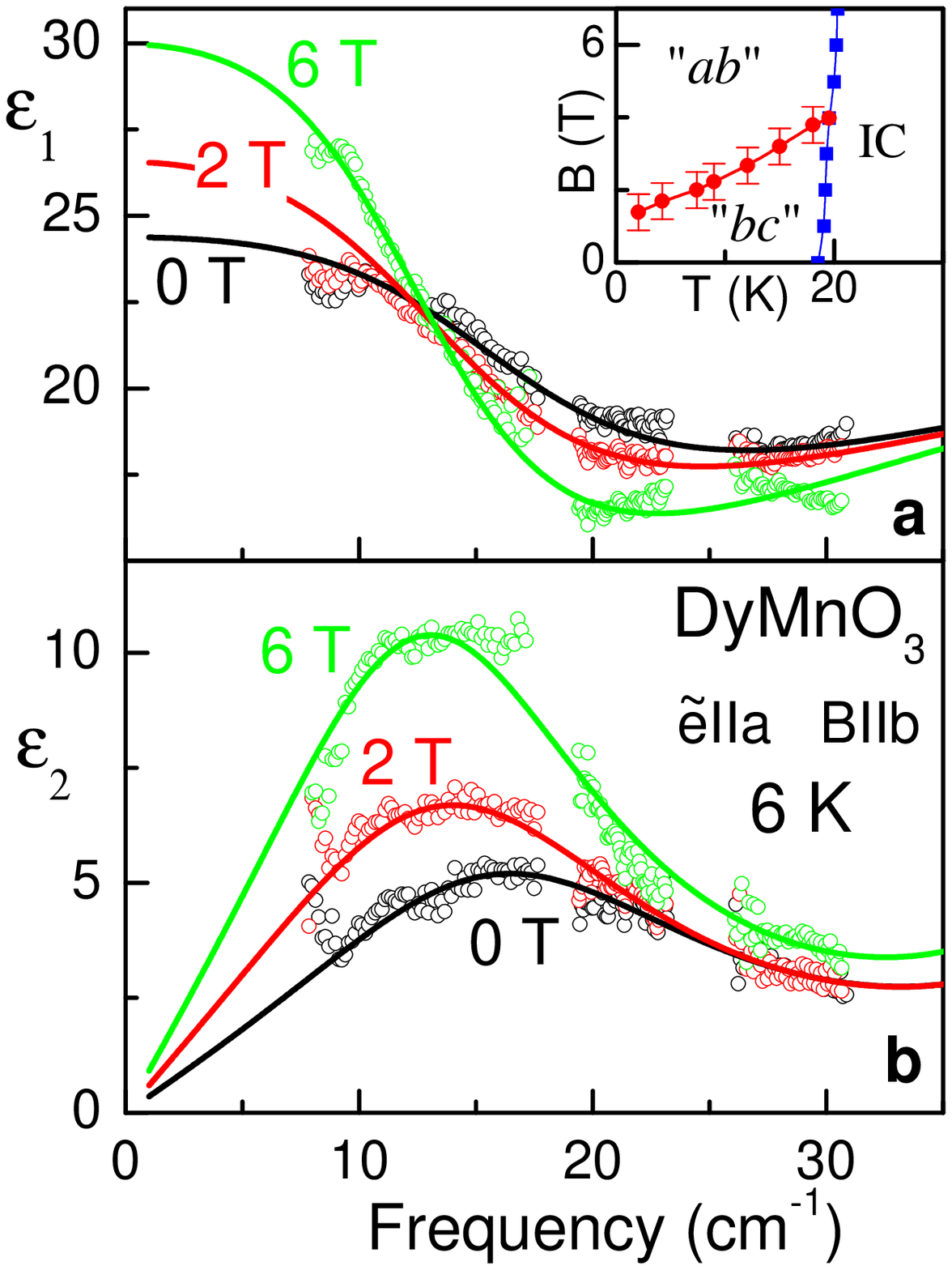}
\caption{Frequency dependence of the real (\textbf{a},\textbf{b})
and imaginary (\textbf{c},\textbf{d}) parts of the dielectric
permittivity of DyMnO$_3$ for $\tilde{e}\|a$ for different external
magnetic fields $B\|b$ and at $T=6$\,K.  The inset shows the (B,T)
phase diagram for static magnetic fields along the $b$ axis.
Notations are the same as in Fig. \ref{fspectra_a}.}
\label{fspectra_b}
\end{figure}

\begin{figure}
\includegraphics[width=0.7\linewidth, clip]{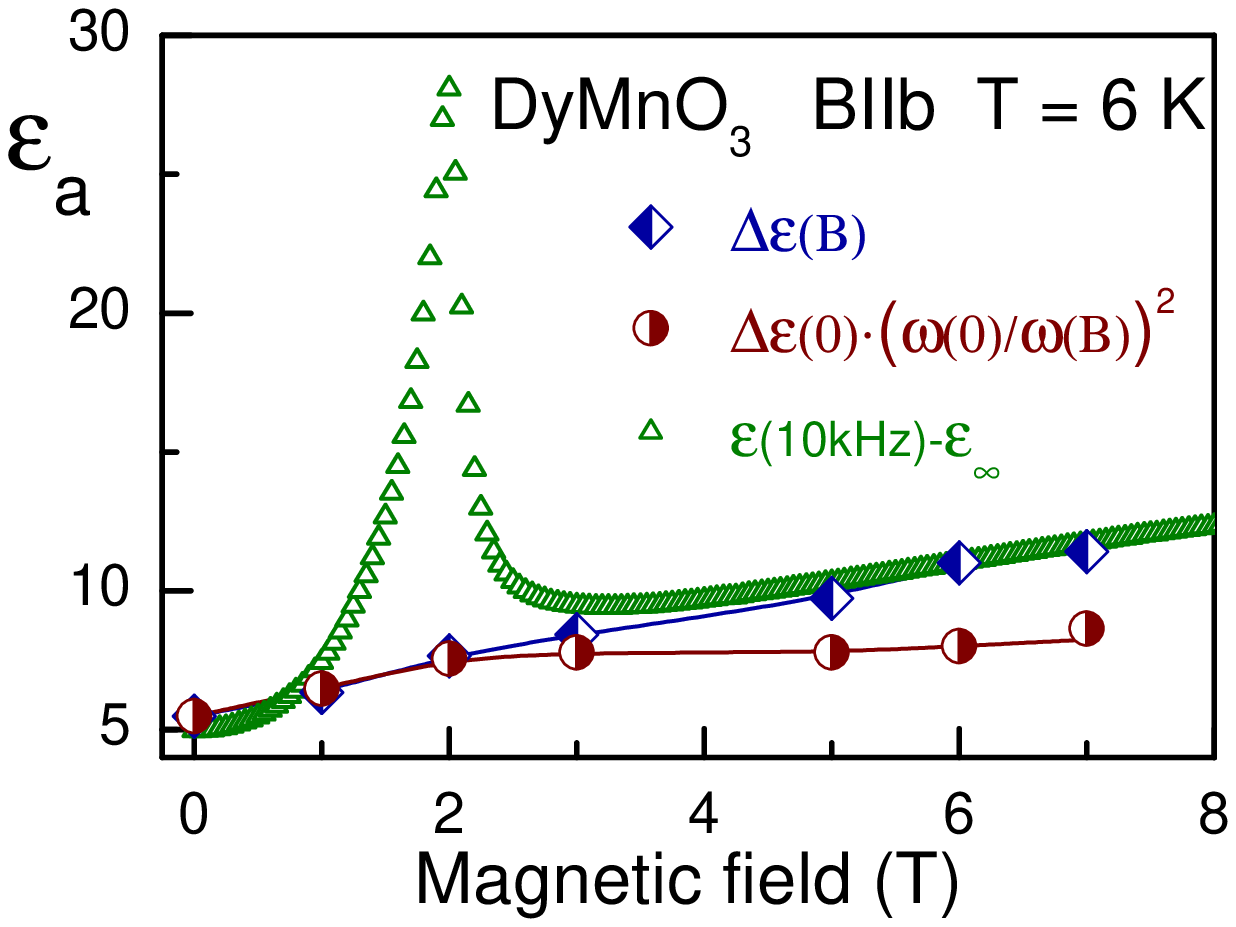}
\caption{Comparison of the static and dynamic properties of
DyMnO$_3$ for $\tilde{e}\|a$ and $B\|b$. Diamonds represent the
measured dielectric contribution of the electromagnon. Circles -
dielectric contribution of the electromagnon as predicted by the LST
relation. Triangles - static dielectric permittivity with a
high-frequency value $\varepsilon_{\infty}=25$ subtracted. }
\label{fpar_b}
\end{figure}

\subsection{$B\|b$}

We turn now to the experimental geometry in which the transition
from the $bc$- to the $ab$ cycloid is achieved by magnetic fields
along the $b$ axis. Although here some hints to a soft mode behavior
could be detected as well, the results turned out to be more
complicated to interpret. Figure~\ref{fspectra_b} reveal the
terahertz spectra of the electromagnon in this geometry. We note at
this point that similar spectra for this field geometry have been
obtained previously in Ref. \cite{kida_prb_2008}. Similar to the
data in Fig.~\ref{fspectra_a}, the spectra in the geometry $B\|b$
reveal an increase of the electromagnon intensity in external
magnetic fields. However, already the comparison of the spectra at
6T and at 2T suggests that the increase of the mode intensity is not
directly correlated with the decrease of the resonance frequency.
Even without exact analysis of the fits one can see that the maxima
in $\varepsilon_2$ for 2T and 6T roughly coincide in spite of
different intensities. Figure \ref{fpar_b} presents the comparison
of the static and dynamic properties for the geometry $B\|b$. The
electromagnon frequency in this geometry softens below 2T and then
remains constant (red circles). Blue diamonds represent the strength
of the electromagnon which increases continuously in the whole range
of the magnetic fields investigated. Contrary to the results for the
$B\|a$ (Fig. \ref{fpar_a}), above the transition to the $ab$ cycloid
at 2T the mode contribution $\Delta\varepsilon$ deviates from the
LST prediction $\Delta\varepsilon \cdot
\frac{\omega(0)^2}{\omega(B)^2}$ (red circles in Fig.~\ref{fpar_b}).
This reflects that the spectral weight of the electromagnon is not
conserved in external fields $B\|b$.

In addition, static dielectric constant as function of magnetic
field is influenced by a second process which leads to the peak-like
feature as documented in Fig. \ref{fpar_b}. Most probably, this
feature has to be described using a different model. In this case we
refer to the recent paper \cite{kagawa_prl_2009} by Kagawa
\textit{et al.} in which the dielectric contribution of the domain
walls in DyMnO$_3$ has been investigated. It has been shown that the
peak in the dielectric constant around $bc$-to-$ab$ phase transition
is due to a domain wall relaxation with characteristic frequency
situated in the radiowave range. We conclude that close to $B=2$T
the main changes in the static permittivity are due to the motion of
the domain walls. However, except for the region close to 2T the
overall increase of the static permittivity corresponds well to the
dielectric contribution of the electromagnon. As a result, the
similarity to the soft mode behavior is quite subtle for the $B \|b$
geometry. Here a better analogy is probably represented by the
motion of the domain walls \cite{kagawa_prl_2009}.  The relaxation
rate of the domain wall motion (Fig. 2 of Ref.
\cite{kagawa_prl_2009}) seem to correlate well with the peak
position of the static dielectric permittivity.

\section{Conclusions}
The behavior of the electromagnon in DyMnO$_3$ in external magnetic
fields has been investigated and compared to the changes of the
static dielectric permittivity. A softening of the electromagnon is
observed for external magnetic fields parallel to the a-axis, which
is accompanied by an increase of the dielectric strength and of the
static permittivity. Our results demonstrate that a soft-mode like
behavior with a substantial increase of the static dielectric
constant can be observed for electromagnons as well. In this case a
magnetoelectric mode becomes soft and an external magnetic field
serves as a driving parameter for this behavior. The observed
similarity reveals further close connection between multiferroics
and classical ferroelectrics. For magnetic fields parallel to the
b-axis no direct correlation between the electromagnon frequency and
the static dielectric permittivity exists. In this case the static
properties seem to be governed by a motion of the domain walls.

\subsection*{Acknowledgements}

We acknowledge fruitful discussion with P. Lunkenheimer. This work
was supported by DFG (Pi 372).

\bibliography{literature}

\end{document}